# Characterizing information leaders in Twitter during COVID-19 pandemic

**David Pastor-Escuredo**[1,2,†], **Carlota Tarazona**[2]

[1]LifeD Lab, Madrid, Spain
[2]Center of Innovation and Technology for Development, Technical University Madrid, Spain
[†]email: david@lifedlab.org https://orcid.org/0000-0002-2662-8686

**ABSTRACT** Information is key during a crisis such as the one produced by the current COVID-19 pandemic as it greatly shapes people's opinion, behavior and their psychology. Infodemic of misinformation is an important secondary crisis associated to the pandemic. Infodemics can amplify the real negative consequences of the pandemic in different dimensions: social, economic and even sanitary. For instance, infodemics can lead to hatred between population groups that fragment the society influencing its response or result in negative habits that help the pandemic propagate. On the contrary, reliable and trustful information along with messages of hope and solidarity can be used to control the pandemic, build safety nets and help promote resilience. We propose the foundation of a framework to characterize leaders in Twitter based on the analysis of the social graph derived from the activity in this social network. Centrality metrics are used to characterize the topology of the network and the nodes as potential leaders. These metrics are compared with the user' popularity metrics managed by Twitter. We then assess the resulting topology of clusters of leaders visually. We propose this tool to be the basis for a system to detect and empower users with a positive influence in the collective behavior of the network and the propagation of information.

## 1 INTRODUCTION

Misinformation and fake news are a recurrent problem of our digital era (Shu et al., 2017; Bakir and McStay, 2018; Lazer et al., 2018). The volume of misinformation and its impact grows during large events, crises and hazards (Allcott and Gentzkow, 2017). When misinformation turns into a systemic pattern it becomes an infodemic (Peters et al., 2020; Zarocostas, 2020). Infodemics are frequent specially in social networks that are distributed systems of information generation and spreading. For this to happen, the content is not the only variable but the structure of the social network and the behavior of relevant people greatly contribute (Zarocostas, 2020).

During a crisis such as the current COVID-19 pandemic, information is key as it greatly shapes people's opinion, behaviour and even their psychological state (Cinelli et al., 2020; Hua and Shaw, 2020; Medford et al., 2020). However, the greater the impact the greater the risk (Vaezi and Javanmard, 2020). It has been acknowledged from the General-Secretary of United Nations that the infodemic of misinformation is an important secondary crisis associated to the pandemic that can amplify the crisis. During a crisis, time is critical, so people need to be informed at the right time (Militello et al., 2007; Greenwood et al., 2017). Furthermore, information during a crisis leads to action, so population needs to be properly informed to act right (Gao et al., 2014). Thus, infodemics can amplify the real negative consequences of the pandemic in different dimensions: social, economic and even sanitary. For instance, infodemics can lead to hatred between population groups (Morales et al., 2015) that fragment the society influencing its response or result in negative habits that help the pandemic propagate. On the contrary, reliable and trustful information along with messages of hope and solidarity can be used to monitor and control the pandemic, create real-time response mechanisms, build safety nets and help promote resilience and antifragility.

To fight misinformation and hate speech, content-based filtering is the most common approach taken (MacAvaney et al., 2019; Pierri and Ceri, 2019; Ghanem et al., 2020; Zarocostas, 2020). The availability of Deep Learning tools makes this task easier and scalable (Ruchansky et al., 2017; Singhania et al., 2017; Popat et al., 2018). Also, positioning in search engines is key to ensure that misinformation does not dominate the most relevant results of the searches. However, in social media, besides content, people's individual behavior and social network properties, dynamics and topology are other relevant factors that determine the spread of information in the different clusters and layers of the society organization skeleton (Iribarren and Moro, 2009; Miritello et al., 2011; Morales et al., 2014).

We propose a framework to characterize leaders in Twitter based on the analysis of the social graph derived from the activity in this social network (Borondo et al., 2015). Centrality metrics are used to identify relevant nodes that are further characterized in terms of users' parameters managed by Twitter (Balkundi and Kilduff, 2006; Goyal et al., 2008; Bodendorf and Kaiser, 2009; Fransen et al., 2015; De Brún and McAuliffe, 2020). Although this tool may be used for surveillance of individuals, we propose it as the basis for a constructive application to detect and empower users with a positive influence in the collective behaviour of the network and the propagation of information (Iakhnis and Badawy, 2019; De Brún and McAuliffe, 2020).

## 2 MATERIALS AND METHODS

### 2.1 Data

Tweets were retrieved using the real-time streaming API of Twitter using a filter of keywords. The keywords were basic terms to retrieve posts related to the pandemic {'coronavirus', 'Coronavirus', '#CoronavirusES', 'coronavirusESP', '#coronavirus', '#Coronavirus', 'covid19', '#covid19', 'Covid19', '#Covid19', 'covid-19', '#covid-19', 'COVID-19', '#COVID-19'}. In total, 500.000 posts were retrieved in the time interval between April 18th and May 4th.

### 2.2 Network construction

Each tweet was analyzed to extract mentioned users, retweeted users, quoted users or replied users. For each post the corresponding nodes were added to an undirected graph as well as a corresponding edge initializing the edge property "flow". If the edge was already created, the "flow" was incremented. The network was completed by adding the property "inverse flow" (1/flow) to each edge. The resulting network featured 107544 nodes and 116855 edges.

### 2.3 Relevance metrics

| | |
|---|---|
| Degree | Radial and volume-based centrality computed from 1-length walks (normalized degree) based on the flow property. This metric measures the number of direct connections that an individual node has to other nodes within a network. |
| Eigenvalue | Radial and volume-based centrality computed from infinite length walks. This metric measures the number of edges per node, and the number of edges of each connected node and so on. |
| Closeness | Radial and length-based centrality that considers the length of the shortest paths of all nodes to the target node based on the flow property. |
| Betweenness | Medial and volume-based centrality that considers the number of shortest paths passing by a target node based on the flow property. This centrality was computed for both directions of the directed graph. |
| Current flow Closeness (cfcloseness) | Radial and length-based centrality based on current flow model using the inverse flow property. This centrality was computed for the largest connected undirected subgraph. This metric is a variant of closeness that evaluates not only the shortest paths, all possible paths. |
| Current flow Betweenness (cfbewteenness) | Medial and volume-based centrality based on current flow model using the inverse flow property. This centrality was computed for the largest connected undirected subgraph. This metric is a variant that evaluates the intermediary position in all the paths between the rest of nodes. |
| Load | The load centrality of a node is the fraction of all shortest paths that pass through that node. Load centrality is slightly different than betweenness. |

**Table 1**. Descriptors table

To compute centrality metrics the network described above was filtered. First, users with a node degree (number of edges connected to the node) less than a given threshold (experimentally set to 3) were removed from the network as well as the edges connected to those nodes. The reason of this filtering was to reduce computation cost as algorithms for centrality metrics have a high computation cost and also removed poorly connected nodes as the network built comes from sparse data (retweets, mentions and quotes). However, it is desirable to minimize the amount of filtering performed to study large scale properties within the network. The resulting network featured 15845 nodes and 26837 edges. Additionally, the network was filtered to be connected which is a requirement for the computation of several of the centrality metrics described below. For this purpose, the subnetworks connected were identified, selecting the largest connected network as the target network for analysis. The resulting network featured 12006 nodes and 25316 edges.

Several centrality metrics were computed: cfbetweenness, betweenness, closeness, cfcloseness, eigenvalue, degree and load. Each of this centrality metric highlights a specific relevance property of a node with regards to the whole flow through the network. Descriptors explanations are summarized in Table 1. Besides the network-based metrics, Twitter user' parameters were collected: followers, following and favorites so the relationships with relevance metrics could be assessed.

## 2.4 Statistics

We applied several statistical tools to characterize users in terms of the relevance metrics. We also implemented visualizations of different variables and the network for a better understanding of leading nodes characterization and topology.

# 3 RESULTS

## 3.1 Descriptors correlation

We compared the relevance in the network derived from the centrality metrics with the user' profile variables of Twitter: number of followers, number of following and retweet count. Figure 1 shows a scatter plots matrix among all variables. Principal diagonal of the figure shows the distribution of each variable which are normally characterized by a high concentration in low values and a very long tail of the distribution. These distributions imply that few nodes concentrate most part of the relevance within the network. More surprisingly, same distributions are observed for Twitter user' parameters such as number of followers or friends (following).

The scatter plots shows that there is no significant correlation between variables except for the pair betweenness and load centralities as it is expected because they have similar definitions. This fact is remarkable as different centrality metrics provide a different perspective of leading nodes within the network and it does not necessarily correlate with the amount of related users, but also in the content dynamics.

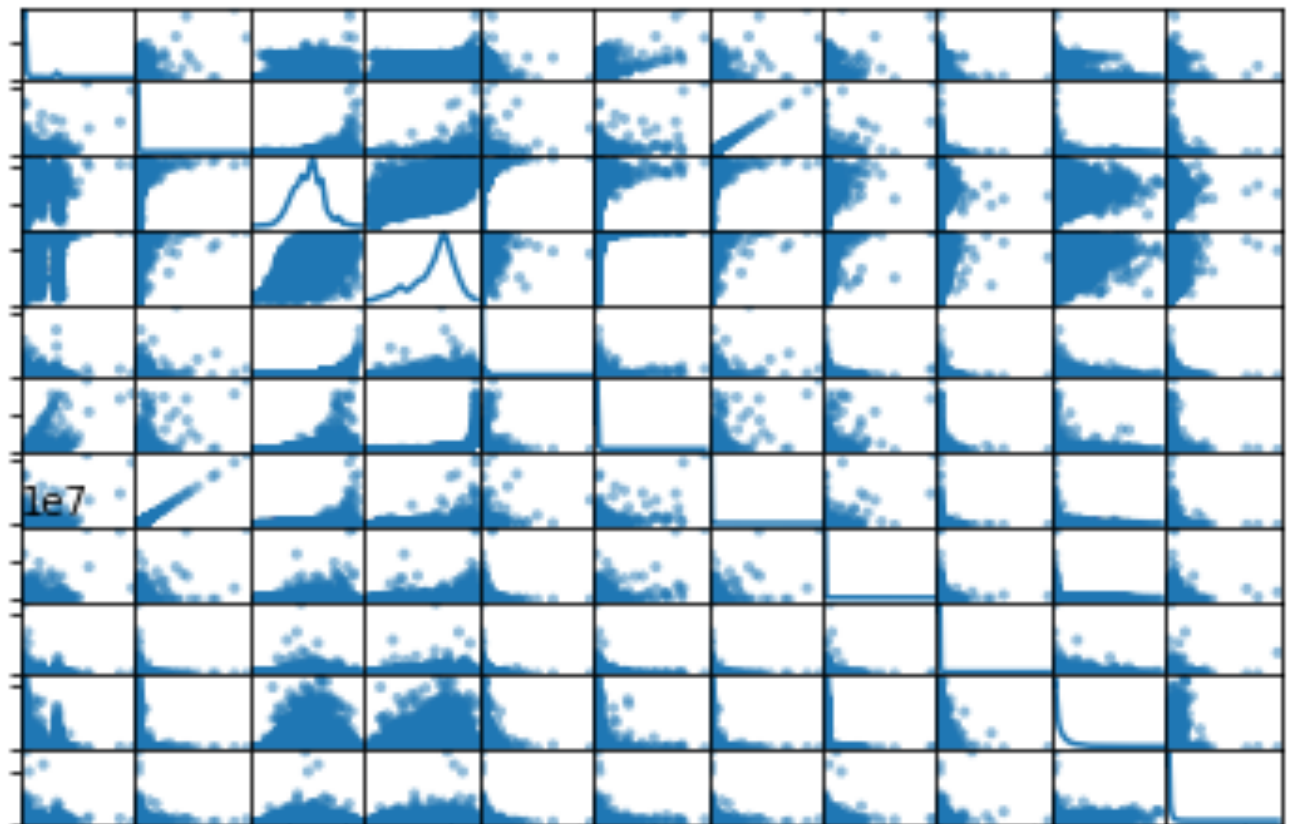

**Figure 1.** Matrix of histograms and scatter plots among all variables. Left to right and from the top to the bottom: cfbetweenness, betweenness, closeness, cfcloseness, eigenvalue, degree, load, followers, following, favorites and status_count.

## 3.2 Ranking

Users were ranked using on variable as the reference. Figure 2 summarizes the values for each descriptor of each leader after being ranked according to the eigenvalue centrality. Figure shows that even within the top ranked leaders there is a very large variability characterized by an exponential distribution for the eigenvalue parameter and very heterogeneous values for other relevance metrics or Twitter popularity metrics (followers, following, favorites and status). The heterogeneous and unequal distribution suggested that a small number of nodes are powerful nodes within the network cumulating most part of the relevance and node connectivity as characterized by the eigenvalue metric. This fact requires further analysis to be interpreted as relevant nodes can be indeed social leaders in society or singular events of the network dynamics within the time window analyzed. Figure 1 and Figure 2 show that relevance may not be directly correlated with popularity or very high in Twitter. This emergent relev

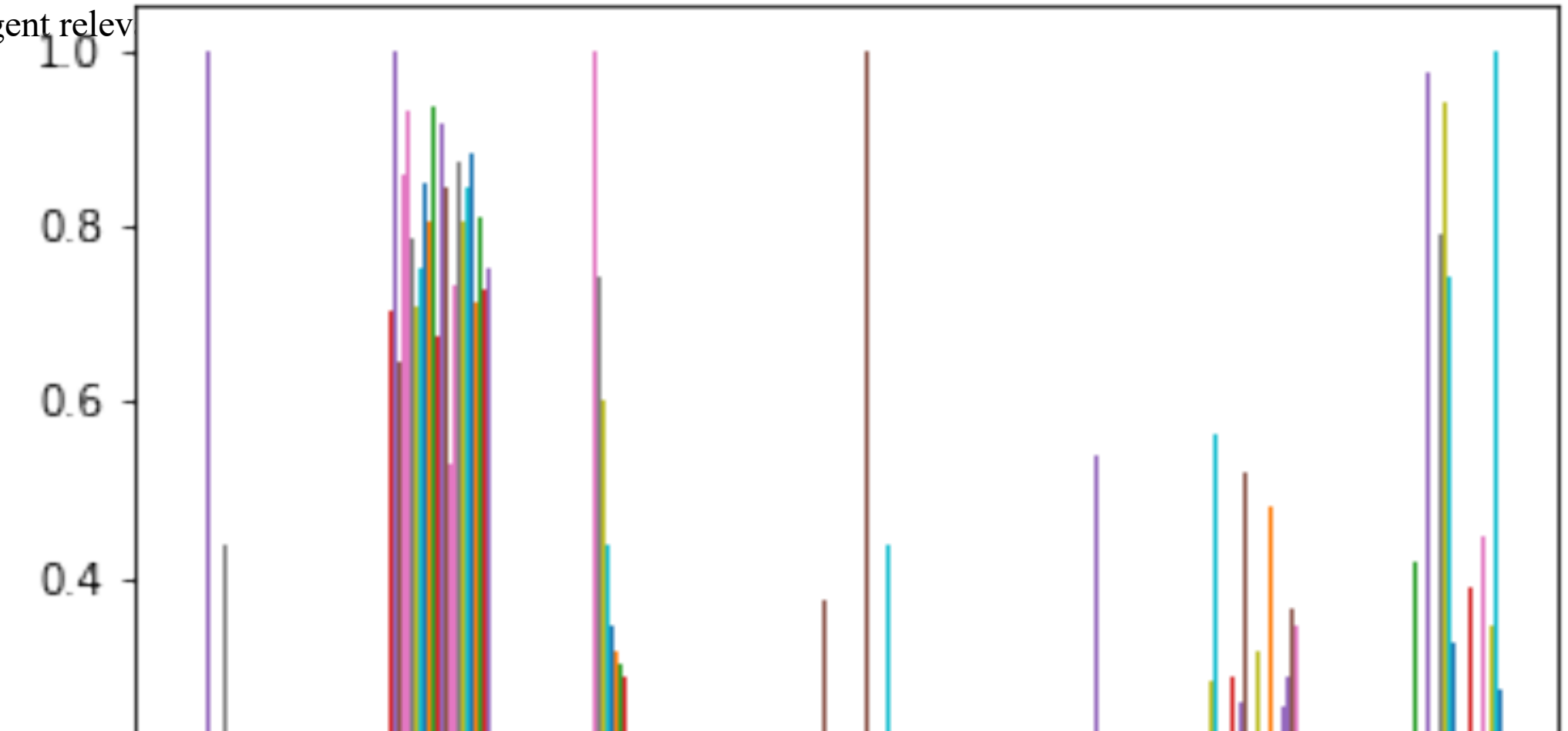

**Figure 2.** Distribution of the ranked users for each descriptor.

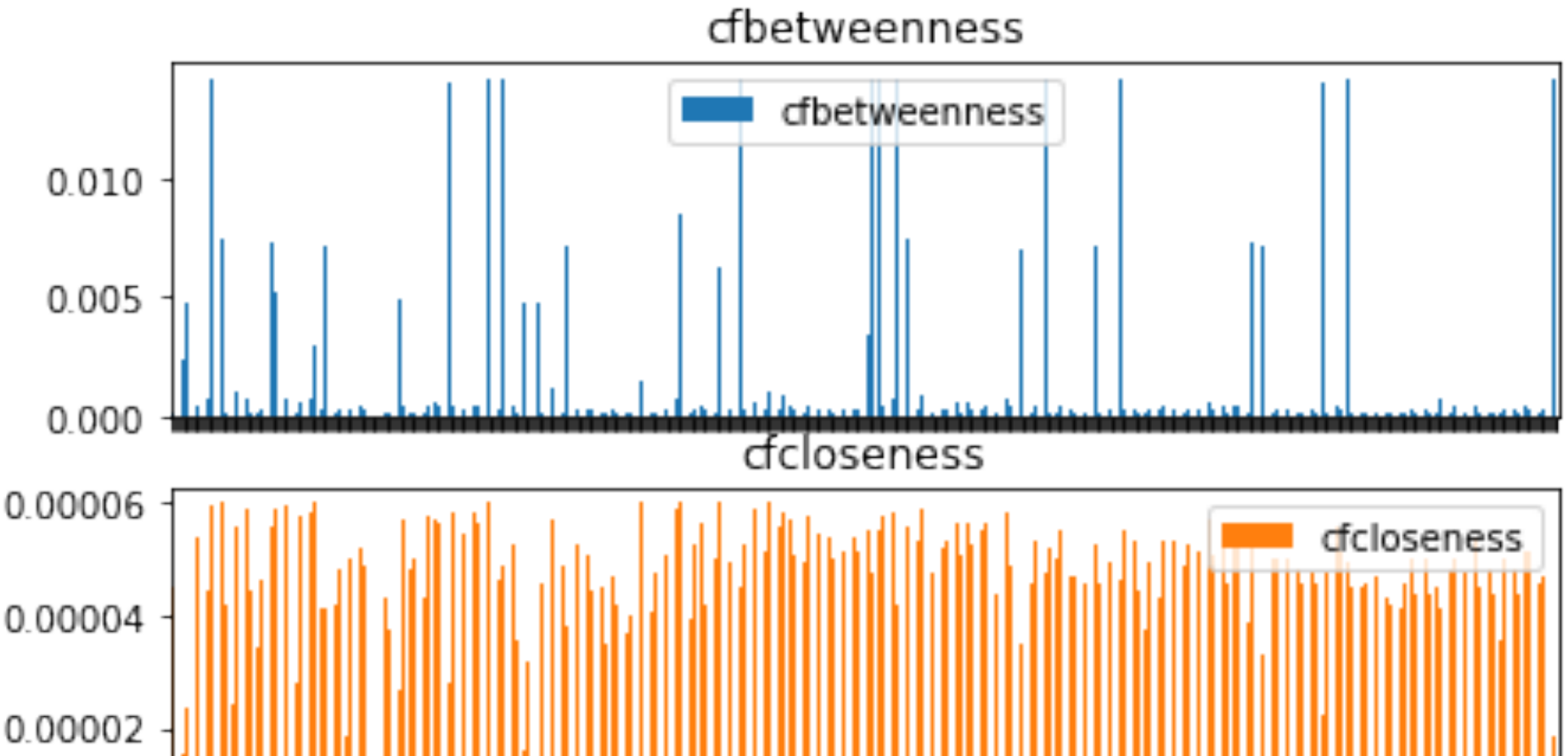

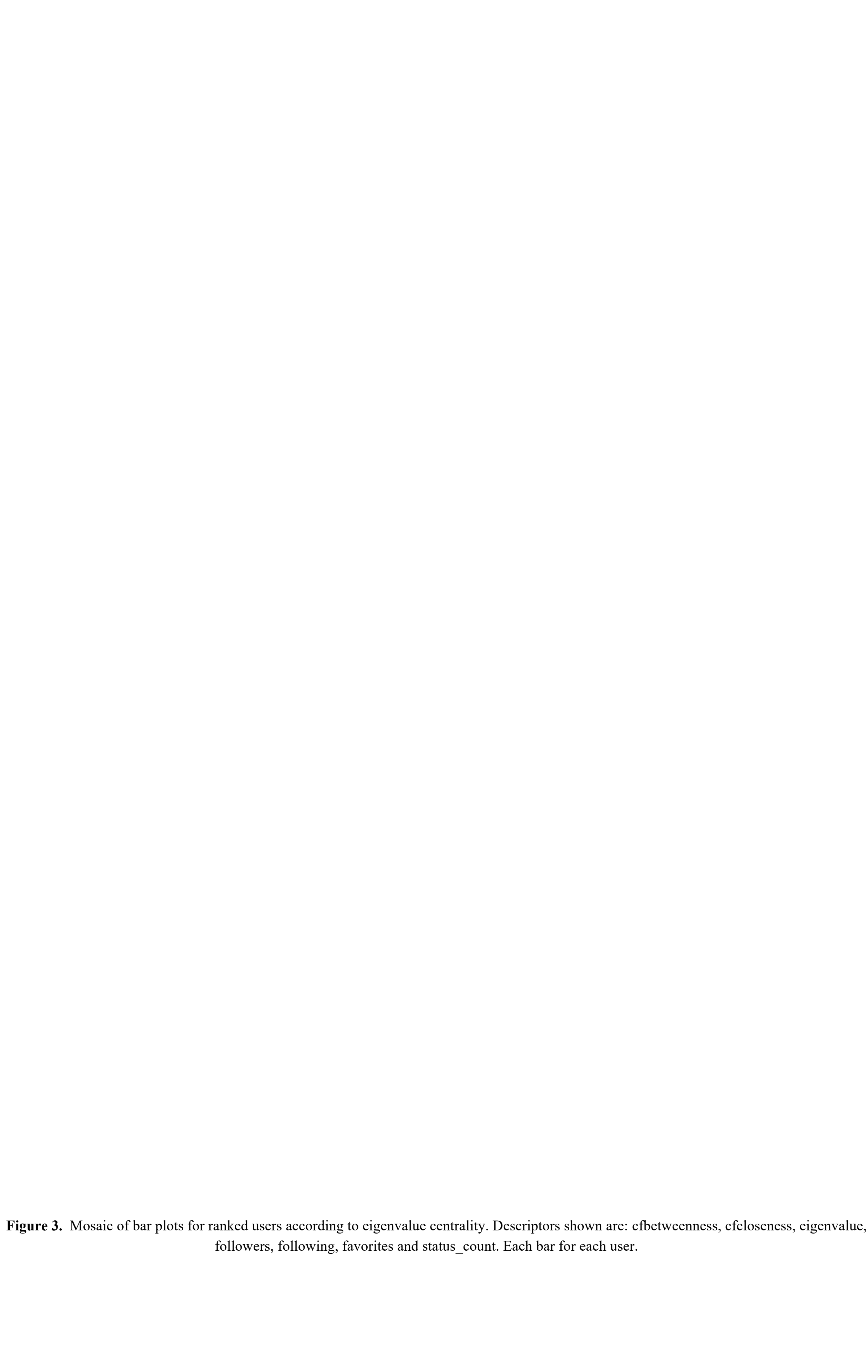

**Figure 3.** Mosaic of bar plots for ranked users according to eigenvalue centrality. Descriptors shown are: cfbetweenness, cfcloseness, eigenvalue, followers, following, favorites and status_count. Each bar for each user.

Figure 3 shows the ranking resulting from using the eigenvalue centrality as the reference. The values were saturated to the percentile 95 of the distribution to improve visualization and avoid the effect of single values with very out of range values. This visualization confirms the lack of correlation between variables and the highly asymmetric distribution of the descriptors. Figure 4 shows the ranking resulting from using current flow betweenness centrality as the reference. In this case, the distribution of this reference variable is smoother and shows a more gradual behavior of leaders. Of note, different centrality metrics lead to different a classification and characterization of nodes (Figure 3 vs Figure 4). Experimental work is required to understand how a specific centrality translates into a different information propagation pattern. For this purpose, we built a dashboard to browse the ranked nodes, their properties and relevance and also the network and its topology.

The occurrence of nodes with centrality values very far away from the distribution average is an important phenomenon when study social leaders. These nodes can play a role of information super-spreaders meaning that a few nodes transmit a lot of information or misinformation to other nodes. This can be appreciated specially in the eigenvalue centrality, whereas the cfbetweenness centrality seems a more stable metric. This asymmetric distribution implies that there are powerful communities highly intra-connected whereas there are fewer nodes that serve as a bridge between communities.

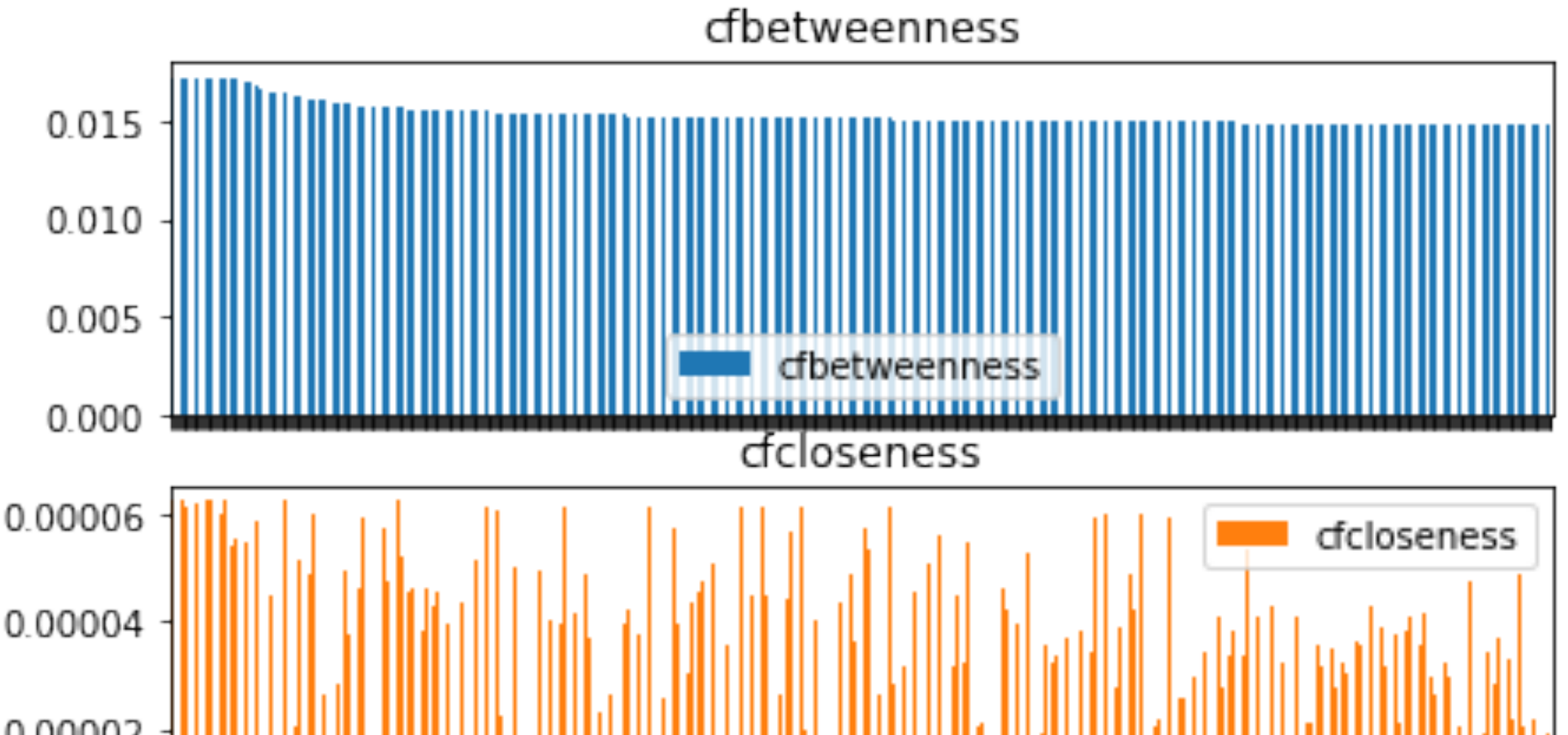

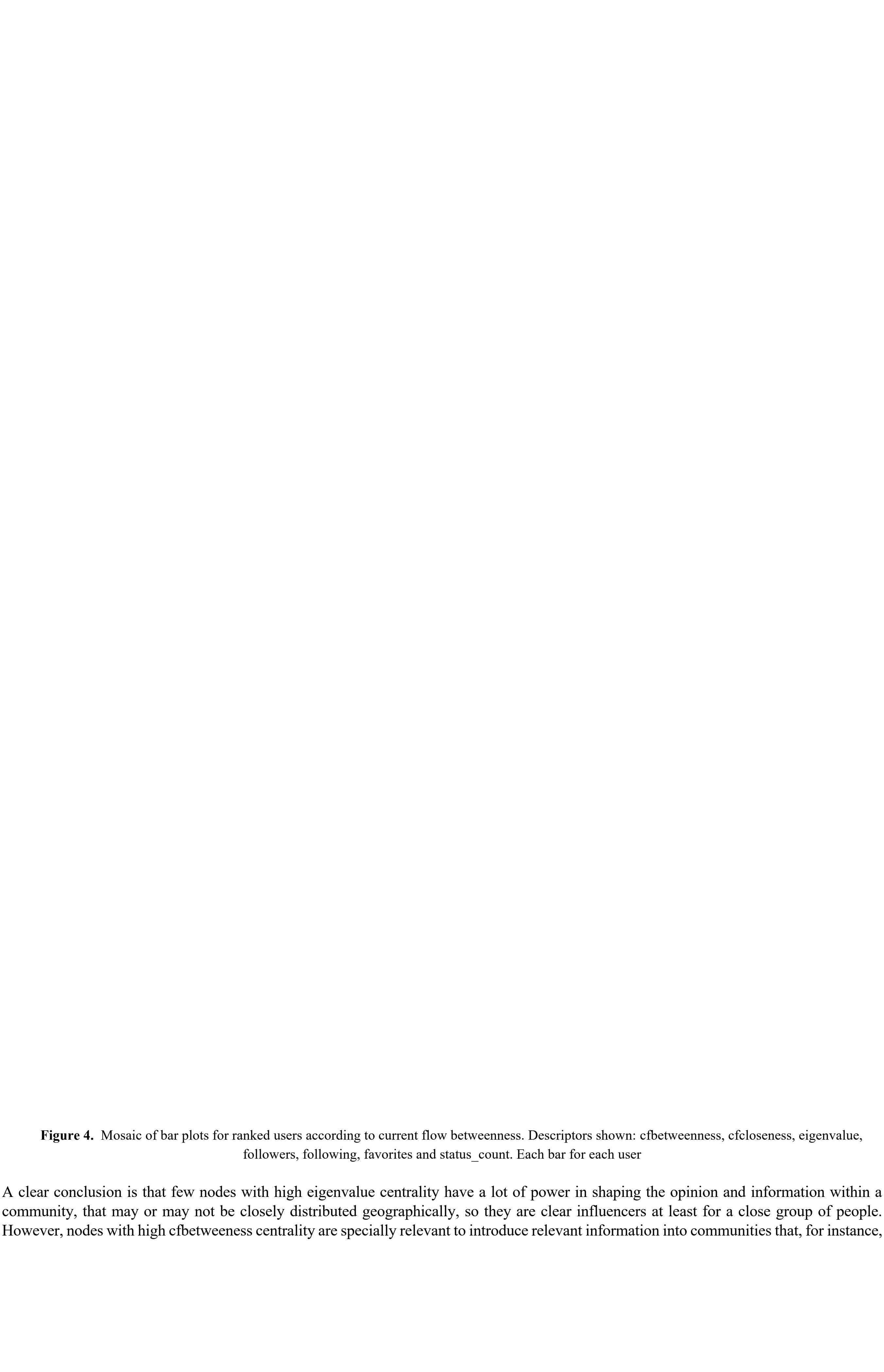

**Figure 4.** Mosaic of bar plots for ranked users according to current flow betweenness. Descriptors shown: cfbetweenness, cfcloseness, eigenvalue, followers, following, favorites and status_count. Each bar for each user

A clear conclusion is that few nodes with high eigenvalue centrality have a lot of power in shaping the opinion and information within a community, that may or may not be closely distributed geographically, so they are clear influencers at least for a close group of people. However, nodes with high cfbetweeness centrality are specially relevant to introduce relevant information into communities that, for instance, may have a negative narrative more indirectly. An open question is that these nodes have the necessary influence within the target community to propagate information and consolidate sentiment. An strategy could be to reinforce nodes with high cfbetweeness and positive activity to be more influential and gain the necessary eigenvalue centrality within communities.

## 3.3 Network

To assess how the nodes with high relevance are distributed we projected the network into graphs by selecting the subgraph of nodes with a certain level of relevance (threshold on the network). The resulting network graphs may not be therefore connected.

The eigenvalue-ranked graph shows high connectivity and very big nodes (see Fig. 5). This is consistent with the definition of eigenvalue centrality that highlights how a node is connected to nodes that are also highly connected. This structure has implications in the reinforcement of specific messages and information within high connected clusters which can act as promoters of solutions, sources of information/misinformation or sentiment or may become lobbies. It is still remarkable that these nodes may not be those with more popularity according to Twitter metrics. It means that for given conversations and topics the dynamic network is not influenced by the popularity of the network. Further analysis is here required to understand the authority that emerges for specific topics based on the content of messages, dynamics of the network and the popularity

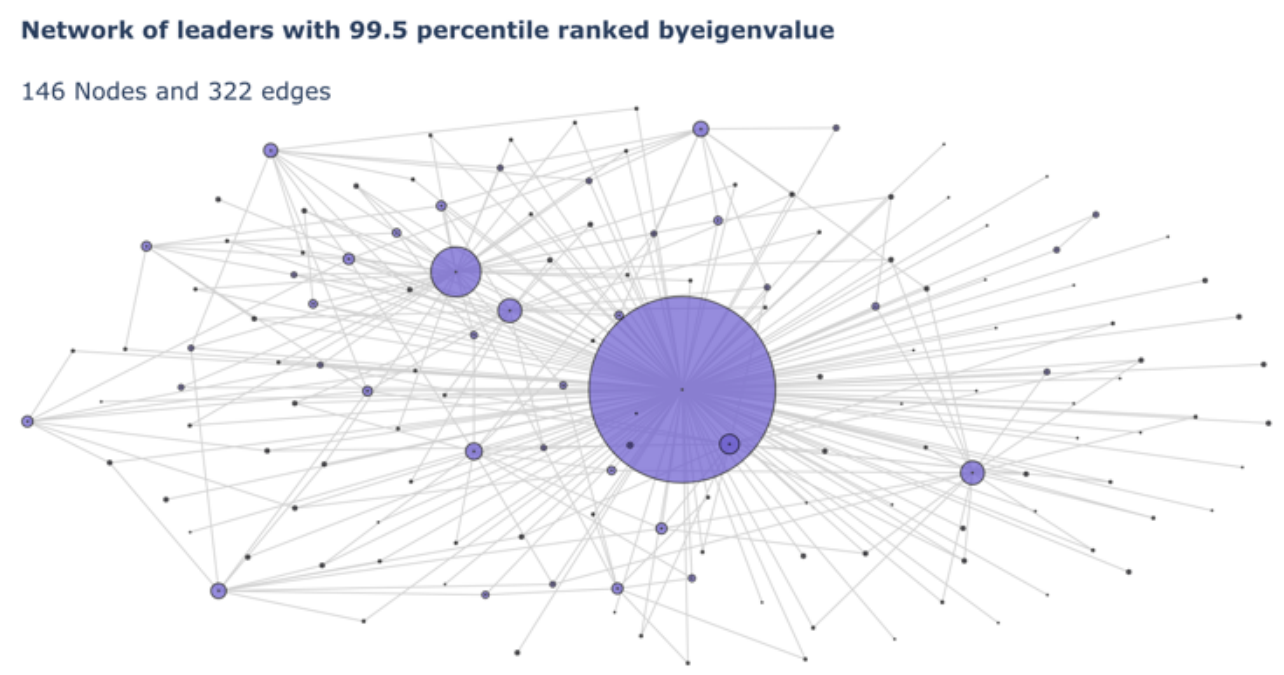

**Figure 5.** Graph of high-eigenvalue users.

The current flow betweenness shows an unconnected graph which is very interesting as decentralized nodes play a key role in transporting information through the network (see Fig. 6). This means that the connectors between communities and groups of opinion are distributed in the network and potentially geographically too. As mentioned, these nodes may not have high popularity or high connectivity (as measured by eigenvalue), but the messages they convey are transmitted across different communities. Further research is required to see the impact of these nodes into the narratives and the propagation of specific sentiments and topics into the communities these nodes interconnect as these nodes have great potential to build larger safety, well-informed and positive nets.

The current flow closeness shows also an unconnected graph which means that the social network is rather homogeneously distributed overall with parallel communities of information that do not necessarily interact with each other (see Fig. 7). These nodes, as the nodes characterized by cfbetweenness have great potential to interconnect communities as they are closer in the network to several communities at the same time. A research question is if the closeness would be sufficient to propagate and consolidate information and sentiment into the target communities.

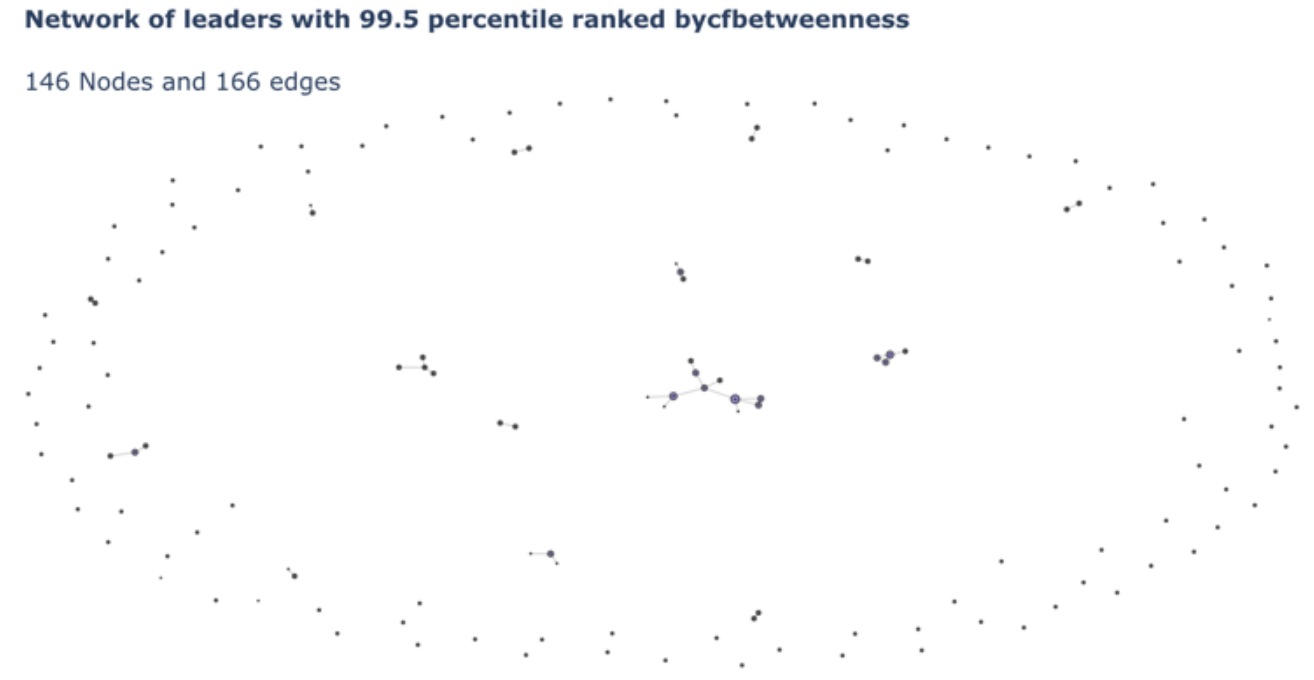

**Figure 6.** Graph of high-current flow betweenness users.

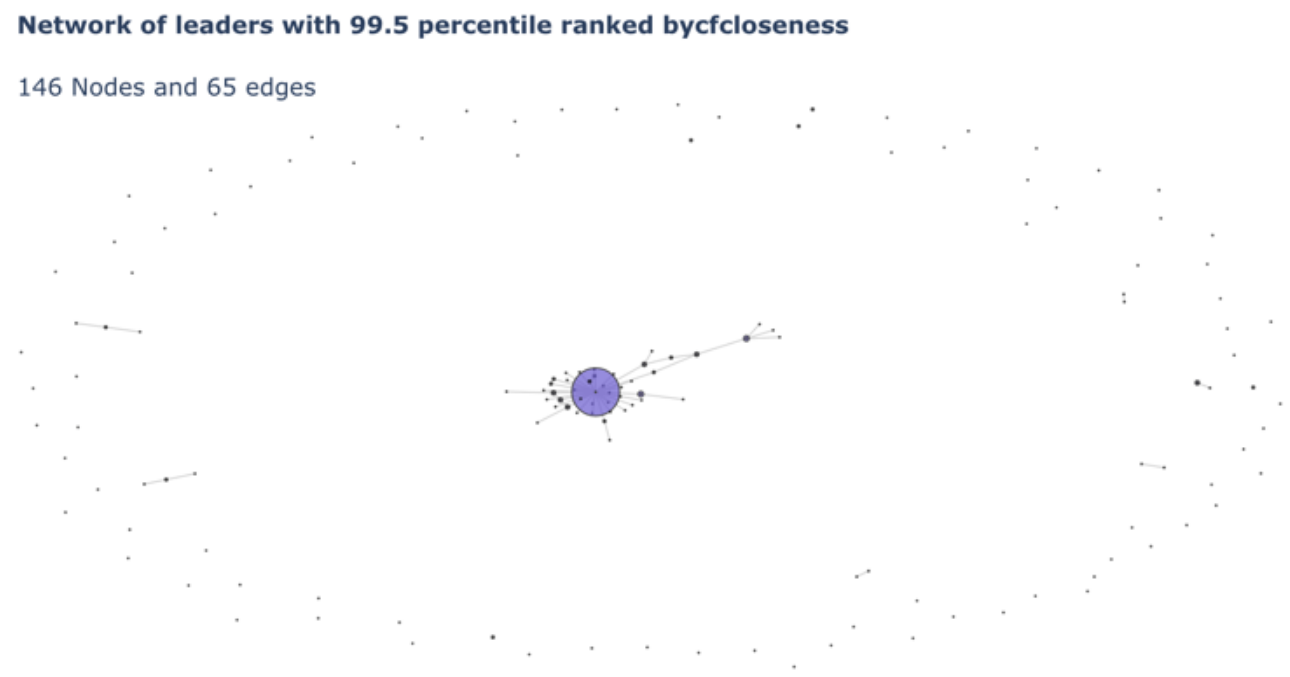

**Figure 7.** Graph of high-current flow closeness users.

By increasing the size of the graph (lowering the thresholds) more clusters can be observed, specially in the eigenvalue-ranked network which is consistent with the previous observations (Fig. 8). A super node may point out to a relevant institution or an anomaly in the network caused by a viral process or topic. The large connectivity of high eigenvalue centrality nodes may be also related to the size of the communities where few communities may be specially large and intra-connected concentrating the flows of information on a specific topic.

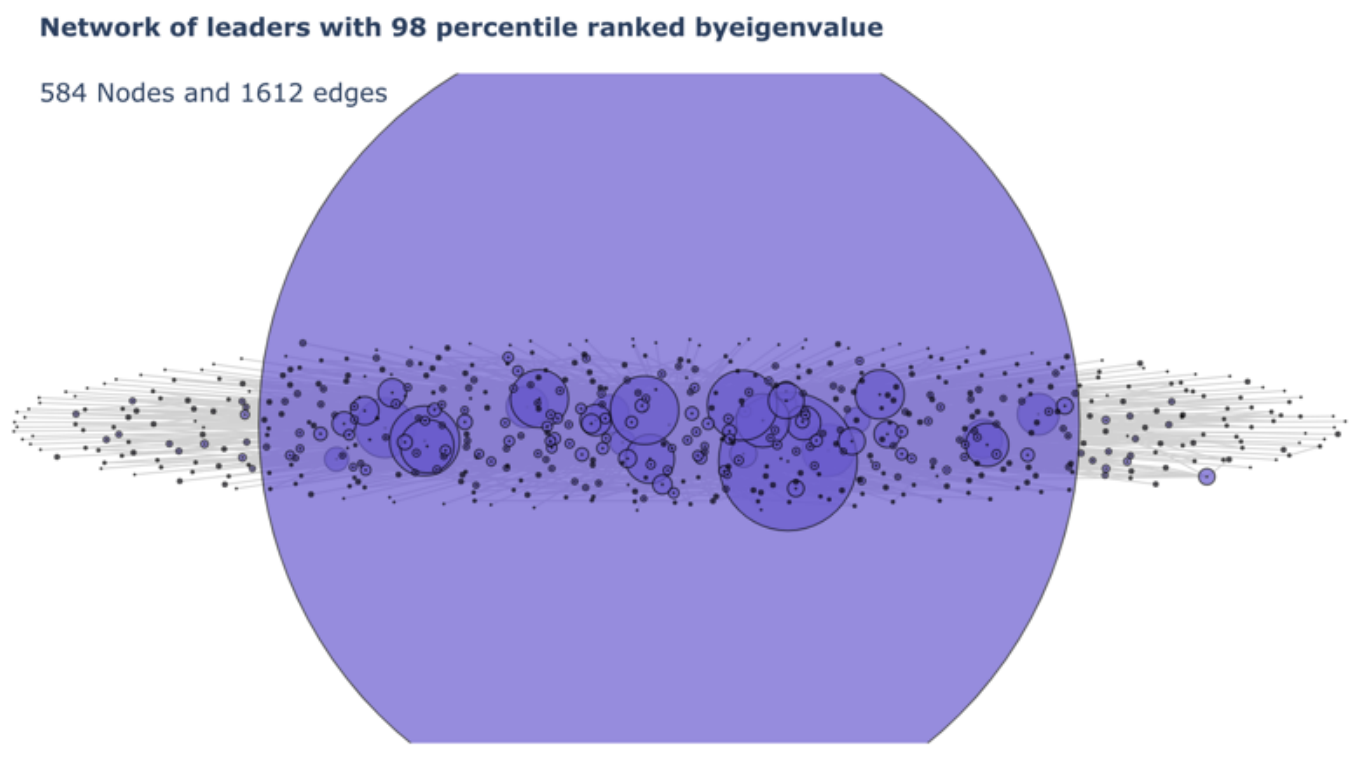

**Figure 8.** Graph of high-eigenvalue users (size 584 nodes).

Some clusters also appear for the current flow betweenness and current flow closeness (see Fig.9 and 10). These clusters may have a key role as highly relevant hubs in establishing bridges between different communities of practice, knowledge or region-determined groups.

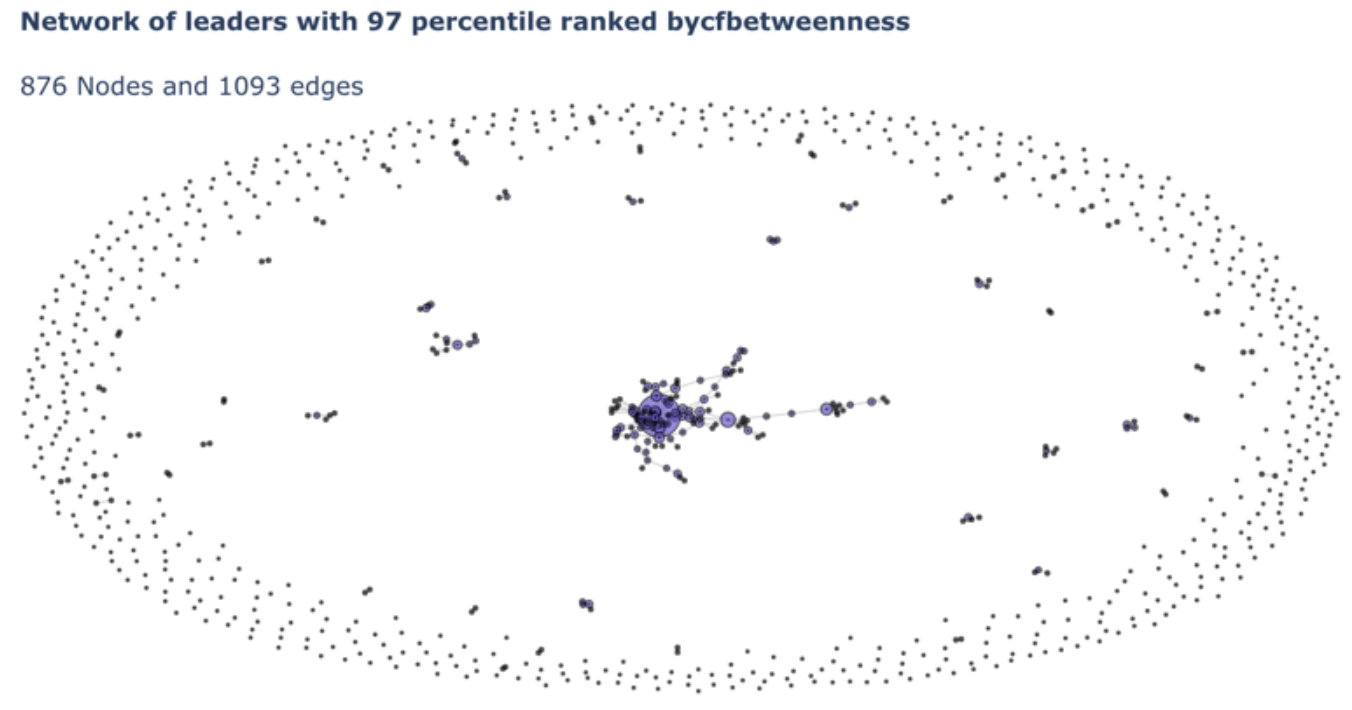

**Figure 9.** Graph of high-current flow betweenness users (size 876 nodes).

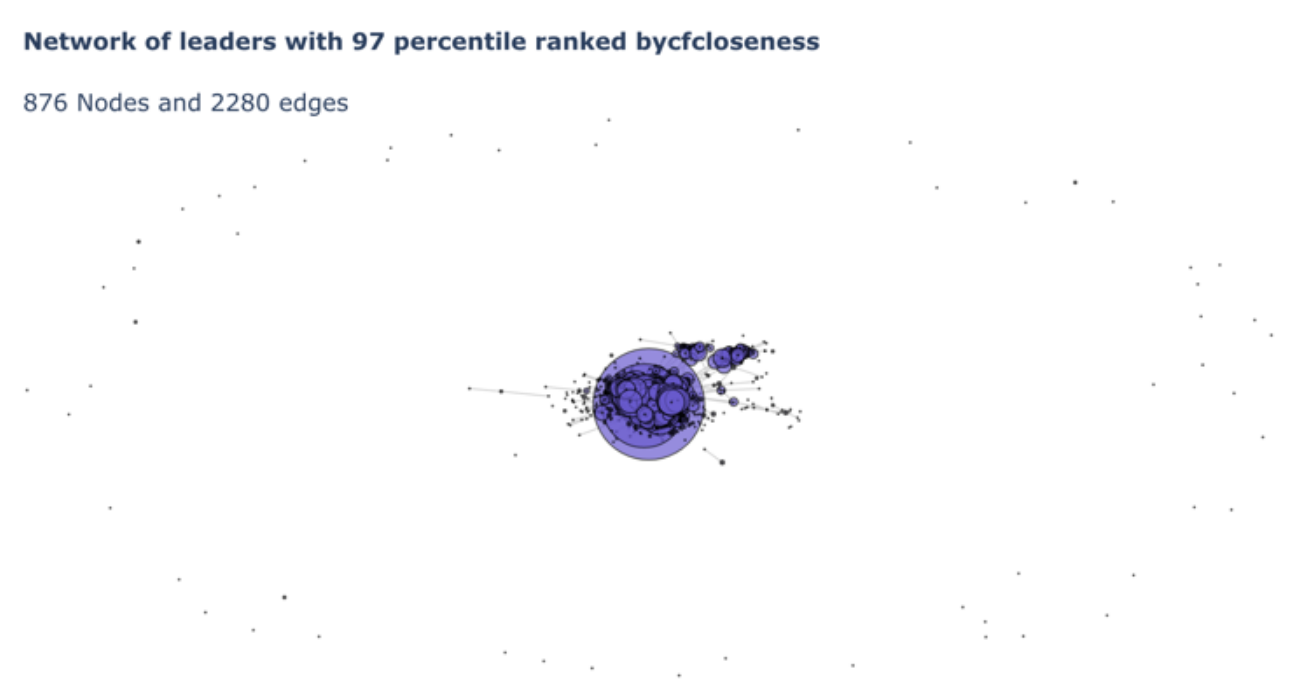

**Figure 10.** Graph of high-current flow closeness users (size 876).

# 4. DISCUSSION

The distributions of the centrality metrics indicate that there are some nodes with massive relevance. As the edges of the network are characterized in terms of flows between users, the relevance should be understood in terms of volume of information between communities or groups that are dynamically connected within a specific topic, in this case information related to COVID-19.

The relevant nodes are topological events within the flow of communication through the network (Morales et al., 2014) that require further contextualization to be interpreted. These nodes can propagate misinformation or make news or messages viral in different ways and with different network length scopes depending on the type of centrality that characterizes them. High eigenvalue nodes will do dense propagation within communities whereas current flow closeness and betweenness will do sparser and widely spread propagation of information. Experimental work is required to optimize the necessary balance between these types of centrality to properly and effectively propagate good information and positive sentiment through the network as an opposition to infodemics.
Further research is required to understand the cause of this massive relevance events, for instance, if it is related to a relevant concept or message or whether it is an emerging event of the network dynamics and topology. Another way to assess these nodes is if they are consistently behaving this way along time or they are a temporal event. Also, it may be necessary to contextualize with the type of content they normally spread to understand their exceptional relevance.

Besides the existence of massive relevance nodes, the quantification and understanding of the distribution of high relevant nodes has a lot of potential applications to spread messages to reach a wide number of users within the network. This is important for inclusiveness of information and to target all types of communities including those that are more vulnerable and can be more affected by infodemics and real impact of the pandemic, both in epidemiological and socio-economic terms. Current flow betweenness particularly seems a good indicator to identify nodes to create a safety net in terms of information and positive messages. The distribution of the nodes could be approached for the general network or for different layers or subnetworks, isolated depending on several factors: type of interaction, type of content or some other behavioral pattern.

We have also developed a first version of an interactive graph visualization to browse the relevance of the network and dynamically investigate how relevant nodes are connected and how specific parts of the graph are ranked to really understand the distribution of the relevance variables.

# 5. CONCLUSION

In this work we leverage network analysis to understand the dynamics of information related to COVID-19. The presented framework and the implemented dashboard are promising tools to fight infodemics and plan strategic information spreading to empower citizens and make more inclusive and responsible use of social networks such as Twitter by building response and smart networks that can help implement values such as inclusiveness, fairness and solidarity that are key in the ongoing crisis that may last a long time.

For a larger impact, experimental work is needed to test, for instance, how a message either positive or negative spreads when started at one of the relevant nodes or close to the relevant nodes. It is also necessary to understand the balance between the behavior of nodes in terms of the different centrality metrics to effectively communicate a narrative. For this purpose, we are working towards integrating a network of concepts and the network of leaders. Understanding the dynamics of narratives and concept spreading is key for a responsible use of social media for building up resilience against crisis.

It is necessary to make a dynamic ethical assessment of the potential applications of this study. Understanding the network can be used to control purposes. However, we consider it is necessary that social media become the basis of pro-active response in terms of conceptual content and information. Digital technologies must play a key role on building up resilience and tackle crisis.

## AVAILABILITY OF DATA AND MATERIAL

The interactive dashboard with the data used for visualizations can be downloaded from https://zenodo.org/record/3996654#.X0LLG9Mza3I. The original Twitter data can be provided upon request.

## CODE AVAILABILITY

The analysis framework code can be provided upon request.

## FUNDING

The authors have not received funding to do this work.

## CONFLICTS OF INTEREST/COMPETING INTERESTS

On behalf of all authors, the corresponding author states that there is no conflict of interest.

## AUTHORS CONTRIBUTIONS

DPE designed and directed the research, designed and performed the analysis, developed the framework, crated figures and wrote the paper. CTL helped develop the framework and created figures.